\documentclass[a4paper]{article}
\usepackage[a4paper,top=3cm,left=3cm,right=2cm,bottom=2cm]{geometry}
\usepackage[affil-it]{authblk}
\usepackage{amsmath}
\usepackage{amsthm}
\usepackage{natbib}
\usepackage{algorithm}
\usepackage{setspace}
\usepackage{fancyhdr}
\usepackage{amssymb}
\usepackage{setspace}
\usepackage[usenames,dvipsnames]{xcolor}

\usepackage{lscape}
\usepackage{subfigure}
\usepackage{graphics}
\usepackage{graphicx}
\usepackage{caption}
\usepackage{csquotes}
\usepackage[noend]{algpseudocode}
\usepackage{longtable}
\usepackage{multirow}
\usepackage{url}
\usepackage[vertfit]{breakurl} 
\usepackage{booktabs}
\usepackage{bigstrut}
\usepackage[shortlabels]{enumitem}
\usepackage{array}
\usepackage{tikz}
\usepackage{rotating}
\usepackage{morefloats}
\usepackage{colortbl}
\usepackage[normalem]{ulem}
\usepackage[titletoc,title]{appendix}

\usetikzlibrary{arrows}
\newcolumntype{L}[1]{>{\raggedright\let\newline\\\arraybackslash\hspace{0pt}}m{#1}}
\newcolumntype{C}[1]{>{\centering\let\newline\\\arraybackslash\hspace{0pt}}m{#1}}
\newcolumntype{R}[1]{>{\raggedleft\let\newline\\\arraybackslash\hspace{0pt}}m{#1}}



\begin{document}
\sloppy
\allowdisplaybreaks
\newcolumntype{H}{>{\setbox0=\hbox\bgroup}c<{\egroup}@{}}




\large
\title{Successful implementation of discrete event simulation: the case of an Italian emergency department}

\author[1,2]{Arthur Kramer\thanks{{\footnotesize email: \texttt{arthur.kramer@ct.ufrn.br}}}}
\author[3]{Clio Dosi}
\author[2]{Manuel Iori}
\author[3]{Matteo Vignoli}

\affil[1]{{\small Departamento de Engenharia de Produ\c{c}\~{a}o\\ Universidade Federal do Rio Grande do Norte, Brazil}}
\affil[2]{{\small Dipartimento di Scienze e Metodi dell'Ingegneria\\ Universit\`{a} degli Studi di Modena e Reggio Emilia, Italy}}
\affil[3]{{\small Dipartimento di Scienze Aziendali\\ Universit\`{a} di Bologna, Italy}}

\date{}

\maketitle

\vspace{-0.75cm}
\begin{center}
 Technical report -- June 2020 \\
\end{center}

\begin{abstract}
	
This paper focuses on the study of a practical management problem faced by a healthcare {\it emergency department} (ED) located in the north of Italy. The objective of our study was to propose organisational changes in the selected ED, which admits approximately 7000 patients per month, aiming at improving key performance indicators related to patient satisfaction, such as the waiting time. Our study is based on a design thinking process that adopts a {\it discrete event simulation} (DES) model as the main tool for proposing changes. We used the DES model to propose and evaluate the impact of different improving scenarios. The model is based on historical data, on the observation of the current ED situation, and information obtained from the ED staff. The results obtained by the DES model have been compared with those related to the existing ED setting, and then validated by the ED managers. Based on the results we obtained, one of the tested scenarios was selected by the ED for implementation.
\end{abstract}

%
\onehalfspace
\section{Introduction}
\label{sec:intro_ED}
{\it Emergency departments} (ED) are complex structures that are prepared to deal with patients with different levels of injuries and diseases, requiring a variety of treatments. In such facilities, even patients with similar characteristics and requirements can follow different paths. Due to these facts, ED attracted the attention of researchers interested in the practical and theoretical aspects related to simulation and optimisation. Concerning practical aspects, overcrowding is a common problem faced by many ED facilities, and it is known as a critical phenomenon  \citep{Trzeciak2003} to be tackled both by the medical and engineering communities \citep{Hoot2008}.

In the last 50 years, the overcrowding issue has been commonly addressed in the research community by the use of simulation modelling tools, albeit with different objectives and techniques (see, e.g., \citealt{Pauletal2010, GulGuneri2015, SalmonEtal2018}). The use of simulation techniques helps in assessing a set of what-if scenarios for organisational decisions taken at different levels \citep{Hulshofetal2012}.
In addition, the interest in managing ED considering the needs of all the involved stakeholders in the systems is growing. This consideration leads to studies on healthcare design. In this context, many works focused on redesigning ED organisational processes to increase efficiency and patient satisfaction and studied how to assess the impact of innovation on health services (\citealt{Herzlinger2006, Madsenetal2006, Prada2008}).

The focus on organisational design and the use of decision support tools (such as simulation modelling) is justified by the aim of reducing the gap between theory and practice, so as to meet the real needs of healthcare organisations (\citealt{Romme2003, Mohrman2007}). Despite the large number of significant studies addressing overcrowding and other related problems in ED, and despite the acceptance of simulation techniques as a relevant contribution instrument to face these problems, nowadays there are still barriers on implementing the results of these studies in practice (see e.g., {\citealt{Foneetal2003, Brailsfordetal2009, Mohiuddine2017, LongAndMeadows2018, Longetal2019}}).

In this work, we investigate a major ED located in the north of Italy, which covers a region with approximately one million inhabitants. The ED mentioned above admits around 80.000 patients per year. We study the integration of a {\it discrete event simulation} (DES) model with a design thinking process aiming at improving some ED {\it key performance indicators} ({KPI}). The DES model acts as a prototyping and learning tool, to help the understanding of the current system and to investigate possible changes to attain performance improvements.

{Preliminary results concerning our case study have been presented at international conferences as \cite{Dosietal2019_iced} and \cite{Dosietal2019_HCSE}. The former focused on an early implementation of our simulation approach, while the latter discussed more theoretical aspects related to the use of simulation in healthcare. This paper concludes our research by presenting extensive results, as well as feedbacks obtained by the ED and the final decisions that have been implemented to reduce overcrowding.}

The remainder of this paper is organised as follows. In Section \ref{sec:Lit_review_ED}, we present a brief literature review of the related works. In Section \ref{sec:meth_ED}, we discuss the methodological aspects developed in this work and Section \ref{sec:case_study} details the case study. Section \ref{sec:experiments_ED} presents the results of our experiments and an analytical analysis of them, while Section \ref{sec:experiments_ED_eval} shows the results from the implementation point of view. Finally, Section \ref{sec:concl_ED} concludes the work.

\section{Literature review}
\label{sec:Lit_review_ED}
\subsection{Simulation tools and implementation limits}
\label{sec:Lit_review_ED_simulation}

In the last 50 years, several studies have approached the ED overcrowding topic recurring to simulation tools. The existence in the literature of many surveys on this topic confirms the high interest of the community on the application of simulation techniques to approach overcrowding and improve efficiency in healthcare units (see, e.g., the surveys by \citealt{GunalPidd2010, Pauletal2010, ABOUELJINANE2013, GulGuneri2015, SalmonEtal2018}).

In the survey by \cite{Pauletal2010}, the authors reviewed works dealing with the problem of ED overcrowding through simulation tools, from $1970$ to $2006$. They identified $43$ relevant articles dealing with this topic, as well as three different categories of ED overcrowding simulation studies: (i) descriptive studies focus on defining overcrowding and looking for causes and effects; (ii) predictive studies focus on measures to predict when an ED would become overcrowded in order to implement a temporary solution using extra resources (such as extra personnel); (iii) intervention oriented studies refer to optimising available resources and processes. Our study is placed in case category (iii). Many other studies approach this category, as, for example, the works by \cite{Wangetal2009}, \cite{ABOHAMAD2013} and \cite{Kuo2016}.
\cite{Wangetal2009} studied an ED located in Lyon, France. The authors modelled the system using a DES model aiming at minimising patients' waiting times. To achieve this goal, two main solutions were proposed and simulated: the first consists considered increasing the doctors’ efficiency and speeding-up some of their activities. The second focuses on creating a dedicated consultation room for low complexity patients. Both solutions provided improvements in the workload of doctors and in the waiting times for consultation by the patients.
In \cite{ABOHAMAD2013}, the authors propose a decision support framework based on the use of simulation modelling, aiming at improving the overall efficiency of the processes. The framework performance has been applied to an ED in a University Hospital in Dublin. \cite{Kuo2016} considered a case study originating from an ED in Honk Kong. They proposed a simulation-optimisation approach to identify the possible intervention points and to evaluate the impact of applying different changing scenarios.

In the literature, several other works involving the application of simulation modelling in healthcare context can be found. Notably, these may concern resource allocation {(\citealt{Ahmed2009, Visintin2017})}, utilisation (\citealt{Santibanez2009}), ambulance location (\citealt{Unluyurt2016}), and layout optimisation (\citealt{Sepulveda1999}), just to cite some. 


Despite the large number of papers dealing with the application of simulation techniques to improve ED efficiency, it is known that there are implementation issues to be overcome in order to have the solution applied in practice. Already in the $1980$s, \cite{Wilson1981} reveals in his study that only $16$ over $200$ computer-simulation projects considered reported successful implementations. Twenty years after, \cite{Foneetal2003} systematically reviewed the use of healthcare simulation models and shed doubt on the value of the implementation `{\it we were unable to reach any conclusions on the value of modelling in health care because the evidence of implementation was so scant. [$\dots$] Further research to assess model implementation is required to assess the value of modelling}'. \cite{Brailsford2007} suggests that nobody has cracked the problem yet and that it is possibly more a social, cultural and educational problem than a technical one. The author evidenced that `{\it Countless projects are carried out by academics and published in academic journals, but these models are not widely taken up by other health providers}'. Still in this sense, \cite{GunalPidd2010} stated that `{\it Even after 25 years of this [Wilson's] review, all these barriers to the successful implementation of simulation still exist to some degree in all domains, including health care}'. It is thus quite remarkable that our simulation study led to a real implementation project supported by the ED.

\subsection{Human-centred design to maximise the chance of implementation}
\label{sec:Lit_review_ED_HCD_implementation}

Nowadays, the organisational design community is given more and more attention to the design thinking process  {(\citealt{Brown2008, Martin2010})}.Design thinking can be seen as a tool that combines the use of innovative methods with the designer's sensibility to satisfy people's needs. It is configured as a model of co-creation and involvement of all the stakeholders in the design (\citealt{Cottam2004}). The idea is that the innovation project has human needs as the central point. Designing around the people involved in the process allows a reduction in the risk related to innovation and increases the chance of implementing the proposed solutions in practice \citep{McCreary2010}.

In many cases, design concepts have been applied to health processes. As discussed, in \cite{Bate2006} and \cite{bessant2009}, designing (or redesigning) healthcare processes from the patients' point of view has been proposed as a key concept to obtain improvements. The main point is that innovation arises in the involvement of patients, doctors, nurses, and process engineers in a shared process based on learning rather than applying best practices.
\cite{BateandRobert2007} investigated the application of the experience-based co-design method in a cancer clinic. While in \cite{bessant2009}, the role of patients in the design process is discussed. In both papers, the authors argued that it is essential to include patients in the co-design method to improve the quality of the offered services. Co-design is also exploited in the study carried out by \cite{Iedemaetal2010}. In \cite{Bevanetal2007}, the authors discuss the use of design thinking paradigms to put together knowledge, research and practice, so as to promote improvements changes on facilities of the national health services in England. \cite{Starninoetal2016} tackled the overcrowding problem in an ED located in Reggio Emilia, Italy. In their design process, they proposed a live prototype putting together staff and patients of the referenced ED, aiming at identifying the needs and the potential intervention point, as well as proposing managerial changes. The implementation of the live prototype indicated a reduction in patient's waiting times and an increase in the overall satisfaction level. \cite{Owadetal2018} and \cite{Sunderetal2020} discussed about the lean methodology to redesign process in study cases in the healthcare context. The former considered an ED in Saudi Arabia, and the latter investigated the integration of lean methodology with design thinking in a mobile hospital in India.

\subsection{{Chance of integrating the two approaches}}
\label{sec:Lit_review_ED_two_approaches}

The literature has presented simulation studies as robust tools to understand, model and decide upon possible statistically relevant interventions. However, the weakness of this approach is in the implementation phase, where stakeholders' reactions and organisational constraints limit the application of the results (see, e.g., \citealt{Mohiuddine2017, LongAndMeadows2018}) .On the other side, the human-centred design community relies on approaches such as human-centred design, participatory design and design thinking to understand needs of stakeholders and decide with them what valuable intervention can be designed and implemented. While this approach has a strong potential, it lacks a strong data-driven decision-making process.

Even though simulation studies and human-centred design present complementary strengths and potentially seem to suggest possible solutions to reciprocal weaknesses, to the author's knowledge, no case studies trying to structurally integrate those two approaches exists. Among the works surveyed by \cite{Pauletal2010}, regarding the use of simulation tools to face overcrowding problems in ED, none of them used simulation as a prototyping and learning tool during the design process. This fact helps in understanding the low rate of simulation results implementation in healthcare and is a strong motivation that supports our research.

Our study develops a methodology that tries to integrate those two approaches and tests it on a case study, considering implementation as a success index of the proposed methodology.

\section{Methodology: simulation integrated into a design thinking process}
\label{sec:meth_ED}

In the design thinking process, designers are asked to answer the needs expressed by all the actors involved in the change, and that is why this approach is defined as a human-centred design (see, e.g., \cite{Brown2011}).

In general, design thinking projects are composed of a $4$-phase iterative exploration cycle that is repeated during the project execution. These phases are: (i) comprehension, (ii) abstraction, (iii) ideation, and (iv) solution.
In phase one, the team is committed on understanding the context, the design challenge and organisation under study. The use of qualitative research tools (e.g., semi-structured interviews and participatory observation) and quantitative research tools (e.g., questionnaires and data analysis) are strongly encouraged. An in-depth analysis of the literature is also recommended aiming at acquiring relevant knowledge for the context and general solution ideas \citep{Romme2003}.
The second phase is where the team builds an abstract model containing what has been understood about the environment under discussion and where the needs of the stakeholders are identified. The main process modelling tools, such as Business Process Model and Notation, system dynamics, agent-based, and discrete event modelling, as well as design tools, such as context map or need map, are used in this step \citep{Vignoli2011}.
The ideation phase is devoted to the generation of the greatest number of possible solutions. The proposition of solutions can be boosted through brainstorming, body-storming and other creative methods. It is also in this phase that concept solutions are selected and built in a prototype.
In the last step in the cycle, the solution phase, the team returns to the field, to verify the emotional, cognitive and functional response of the organisation to the prototypes made, and then restart the cycle. Each phase is connected with the previous and the following phases. The iteration between the phases is highly recommended as it is expected that the team members first assume a divergent exploratory attitude, generating a wide range of concepts, and later converge into a smaller set of solutions. However, the phases do not need to be carried out in the presented order.

In our work, the four design thinking phases have been tested in a major Italian ED to reduce patients' waiting times and improve employees' satisfaction. A DES model has been developed to support the design thinking process. In the abstraction phase, the DES model has been used as a tool to reproduce the current ED setting and to evaluate what has been collected and analysed in the comprehension phase. Also, the DES model had considerable importance in the design thinking ideation phase, where solutions were proposed, tested, and chosen.

\section{Case study}
\label{sec:case_study}

In this study, we investigate an ED located in the north of Italy that covers a region with more than one million inhabitants and admits more than $80.000$ patients per year. Given the general dissatisfaction of ED employees (doctors, nurses, and aid nurses) and conflicts among professionals, the hospital top management and the head physician asked us for support. The project lasted $18$ months. The ED design aimed to improve the actual processes, to find possible ways to improve the ED system in general and ED professionals working habits in particular so that professionals could be supported in their everyday routines. The top management asked for the design of a solution that had to be implementable. It is important to note that the ED under study was renowned as a conservative and hard-to-manage organisation. In the last $15$ years, most of the interventions proposed by different actors failed to be implemented. We approached the context with a design thinking process, sided with simulation studies, and tested it on a real case study. We created an ad-hoc group with professionals of the ED department that were involved in the design process and the decision making. The group met once every ten days, and the hospital top management was involved once every three months.

Considering this context, in the design thinking comprehension phase, we collected key information by using different methods. Initially, the ED staff provided a database containing historical data. In addition to that, essential information such as patient flows through the ED and patient and ED staff needs were collected by interviews and direct observation. The ED description and the data collection and analyses performed are detailed in Sections  \ref{sec:ED_description} and \ref{sec:data_ED}.
In the abstraction phase, a DES model has been developed. The model considers the data and information collected during the comprehension phase to replicate the ED current setting. Initially, the DES model has been set-up with the aim of better understanding the system under study and identifying the problems to be tackled. Firstly, the current system setting has been modelled, the model validated, and the problems identified. In a second moment, already in the ideation and solution phases, the model has been used as a prototyping tool for proposing and evaluating what-if improvement scenarios. The main steps concerning the abstraction phase are presented in Section \ref{subsec:validation_ED}, while those concerning the ideation and solution phases are discussed in Sections \ref{sec:scenarios_ED} and \ref{sec:experiments_ED}.

\subsection{ED description: Patient flow}
\label{sec:ED_description}

\begin{figure}[!htpb]
	\centering
	\includegraphics[width=0.79\textwidth]{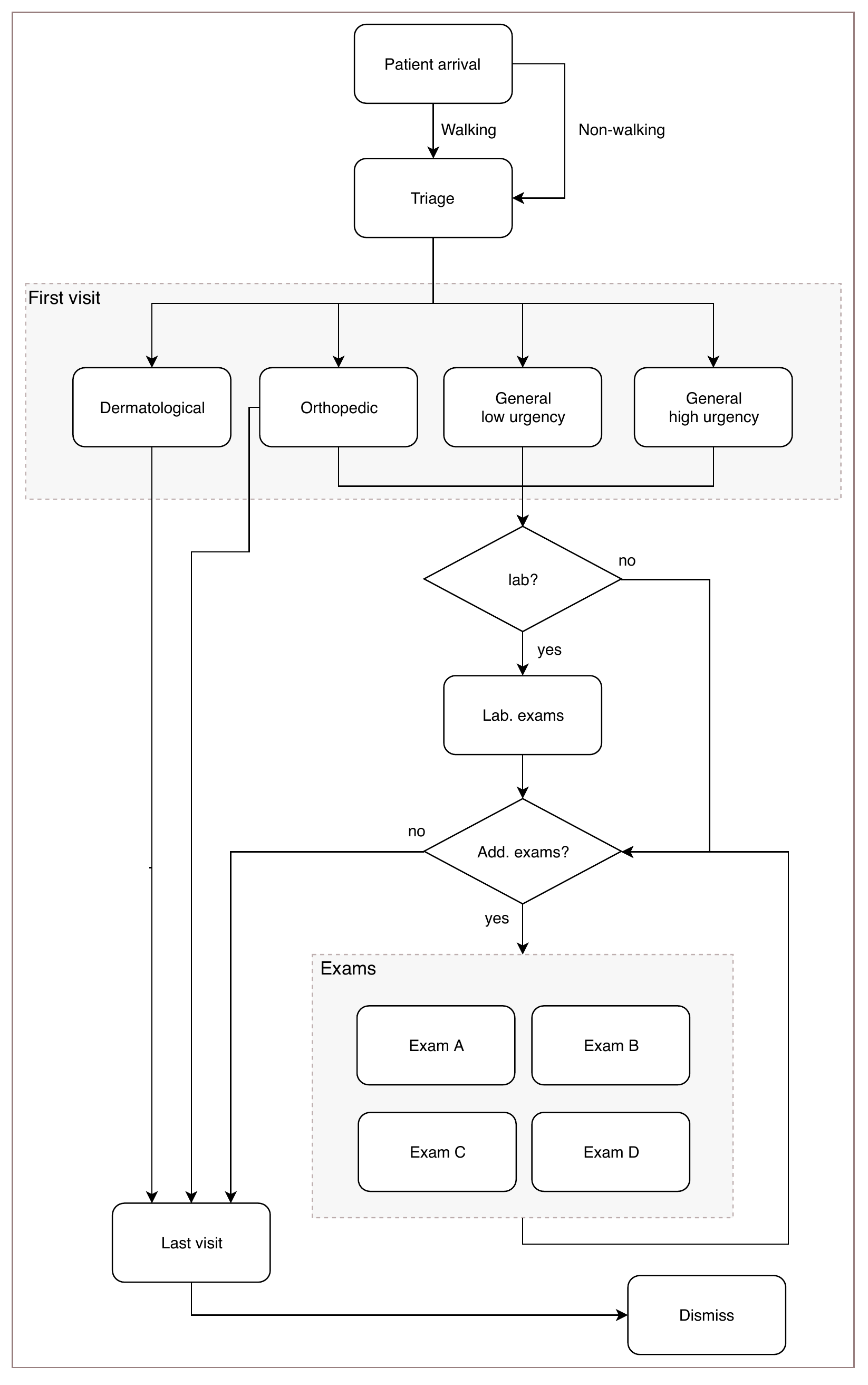}
	\caption{ED flowchart}
	\label{fig:flowchart}
\end{figure}

Before presenting the proposed DES model, it is important to describe the ED structure under study. Since the ED layout is closely related to the patient flows throughout the ED, in the following, we detail the main paths that a patient can follow inside the system. The flow of patients in the ED has been identified by analysing historical data provided by the ED managers and by in loco observation. The identified ED flow of patients is depicted in Figure \ref{fig:flowchart}.
Initially, the patients arrive at the ED following two possible paths. The first one, referred to as ``walking'', concerns mainly the non-urgent patients, while the ``non-walking'' one is mostly used by urgent patients arriving at the ED by ambulance. Both paths are followed by the triage process, which is responsible for identifying the patients' needs and for assigning them to an urgency category. The ED under study considers four urgency levels classified, from the least to the most urgent, as (i) white, (ii) green, (iii) yellow, and (iv) red. In practice, patients classified with the red code are immediately directed to the high urgency general treatment room. The patients with other urgency codes are directed to the waiting room where they wait for being treated. Patients who need a dermatological or an orthopaedic treatment are served exclusively at the emergency dermatological and orthopaedic rooms, respectively. Patients requiring a general visit can be served at the low or high urgency general rooms. Those with red or yellow urgency code are treated exclusively at a high urgency room. Those with white and green codes are likely to be handled by the low urgency room, but they can be treated on a high urgency room if one is idle.

Once the first visit is done, the physicians can require further examinations, including, for example, laboratory analysis and x-ray exams. If laboratory analysis is requested to be performed, then it should be done before all other further required exams, because the laboratory results are needed for those additional exams. Thus, there exists a precedence relation between the laboratory and the other additional exams. The laboratory exams are performed in a dedicated area located in a building close to the ED main building. Due to this fact, the test tubes are transported from the main building to the laboratory building. This transportation is performed every half hour. Thus, the waiting time for transportation and the transportation time itself impacts directly on the total laboratory examination time. The actual laboratory capacity is so high that it can be considered unlimited in practice. Among the additional exams, it is important to highlight the x-ray area as it has a high demand, differently from the other ones that have a very large capacity compared to their demand. When all additional examinations are done, the patient goes back to the waiting room. At this time, he/she is eligible to attend the last visit, which is performed by the same doctor who performed the first visit. Finally, after this last visit, the patient is dismissed and leaves the ED.

\subsection{Data analysis}
\label{sec:data_ED}

Alongside with the patient flows, some additional data were required as input data to set-up our simulation model: patient arrival rates, patient urgency distributions, resources availability, schedules, service times distribution, and queue rules, among others. To retrieve this information, we applied two different techniques: (i) data collection and analysis, and (ii) quantitative time and motion observation. The former allows obtaining the required inputs by analysing historical data, i.e., studying what happened in the past. The latter is usually adopted when historical data is scarce or does not permit obtaining the main inputs needed. In our case, a consistent database containing nine months of quantitative data is available, but some information related to the service times could not be straightly obtained.

\begin{figure}[!htpb]
	\centering
	\begin{tikzpicture}[thick, scale=0.6, every node/.style={transform shape}, >=stealth', dot/.style = {draw, fill = white, circle, inner sep = 0pt, minimum size = 4pt}]
	\coordinate (O) at (0,0);
	\draw[-] (-0.3,0) -- (14,0) coordinate[label = {below:time}] (xmax);
	\draw[-] (-0.3,3) -- (14,3) coordinate[label = {below:}] (x'max);
	\draw[-] (.7,-0.5) -- (.7,6) coordinate[label = {right:}] (ymax);
	\node[rotate=90,anchor=north] at (0.0, 1.5){{Low urgency}};
	\node[rotate=90,anchor=north] at (0.0, 4.5){{High urgency}};
	\node[] at (2, -0.7 ){{8:00}};
	\node[] at (7.5, -0.7 ){{20:00}};
	\node[] at (13, -0.7 ){{8:00}};
	\draw[-, thin] (2,-0.2) -- (2, 0.2);
	\draw[-, thin] (7.5,-0.2) -- (7.5, 0.2);
	\draw[-, thin] (13,-0.2) -- (13, 0.2);
	\draw[-, thin] (2, 0.75) -- (7.5, 0.75);
	\node[] at (4.75, 1){Team A};
	\draw[-, thin] (2, 0.55) -- (2, 0.95);
	\draw[-, thin] (7.5, 0.55) -- (7.5, 0.95);
	\draw[-, thin] (2, 2.25) -- (7.5, 2.25);
	\node[] at (4.75, 2.5){Team B};
	\draw[-, thin] (2, 2.05) -- (2, 2.45);
	\draw[-, thin] (7.5, 2.05) -- (7.5, 2.45);
	\draw[-, thin] (2, 3.75) -- (7.5, 3.75);
	\node[] at (4.75, 4){Team C};
	\draw[-, thin] (2, 3.55) -- (2, 3.95);
	\draw[-, thin] (7.5, 3.55) -- (7.5, 3.95);
	\draw[-, thin] (2, 5.25) -- (7.5, 5.25);
	\node[] at (4.75, 5.5){Team D};
	\draw[-, thin] (2, 5.05) -- (2, 5.45);
	\draw[-, thin] (7.5, 5.05) -- (7.5, 5.45);
	\draw[-, thin] (7.5, 3.75) -- (13, 3.75);
	\node[] at (10.25, 4){Team E};
	\draw[-, thin] (7.5, 3.55) -- (7.5, 3.95);
	\draw[-, thin] (13, 3.55) -- (13, 3.95);
	\draw[-, thin] (7.5, 5.25) -- (13, 5.25);
	\node[] at (10.25, 5.5){Team F};
	\draw[-, thin] (7.5, 5.05) -- (7.5, 5.45);
	\draw[-, thin] (13, 5.05) -- (13, 5.45);
	\end{tikzpicture}
	\caption{Teams shift}
	\label{fig:shift}
\end{figure}
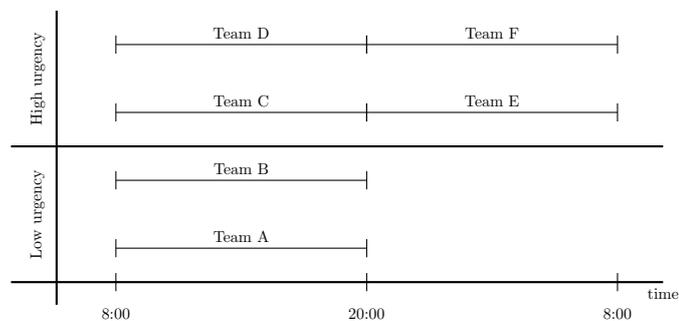

Information about personnel work shifts and resource availability was provided by the ED staff, while the ones concerning queue rules/priorities currently adopted by the organisation have been retrieved during the observation step and by interviewing experts. The ED has six working teams available per day to serve the incoming patients in the current setting. These teams are mainly dedicated to the general first and last visits because the dermatological, orthopaedic and additional examination areas have their dedicated teams. The teams are composed mainly by doctors and nurses and are divided into two groups, one for the low and the other for the high-intensity areas. Each day, four teams work on the shift from 8:00 to 20:00 (two at the low and other two at the high urgency area), and other two teams work from 20:00 to 8:00 dedicated to both urgency areas, as depicted in Figure \ref{fig:shift}.

Furthermore, consistent quantitative information regarding the distributions of patients' arrival rates, urgency and exam requirements have been directly extracted from the database, as well as the service times of additional exams (such as laboratory and x-ray). Information about some service times (such as for the general visits) that could not be directly obtained from the database was obtained by {\it in loco} observation and by interviewing ED’s staff.

\begin{figure}[htb]
	\centering
	\includegraphics[width=0.75\textwidth]{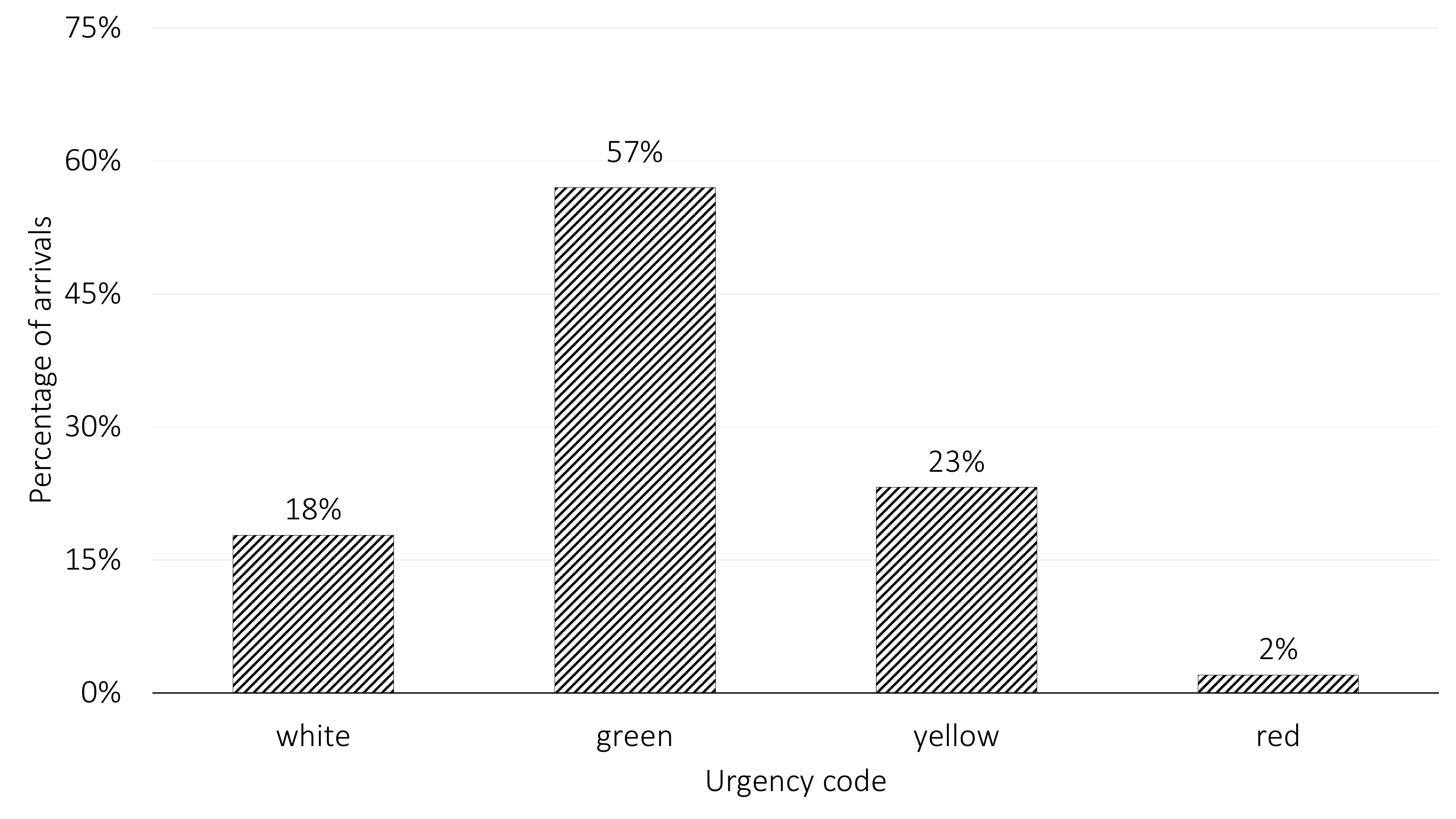}
	\caption{Urgency code distribution}
	\label{fig:urgency_dist}
\end{figure}

\begin{figure}[htb]
	\centering
	\includegraphics[width=0.75\textwidth]{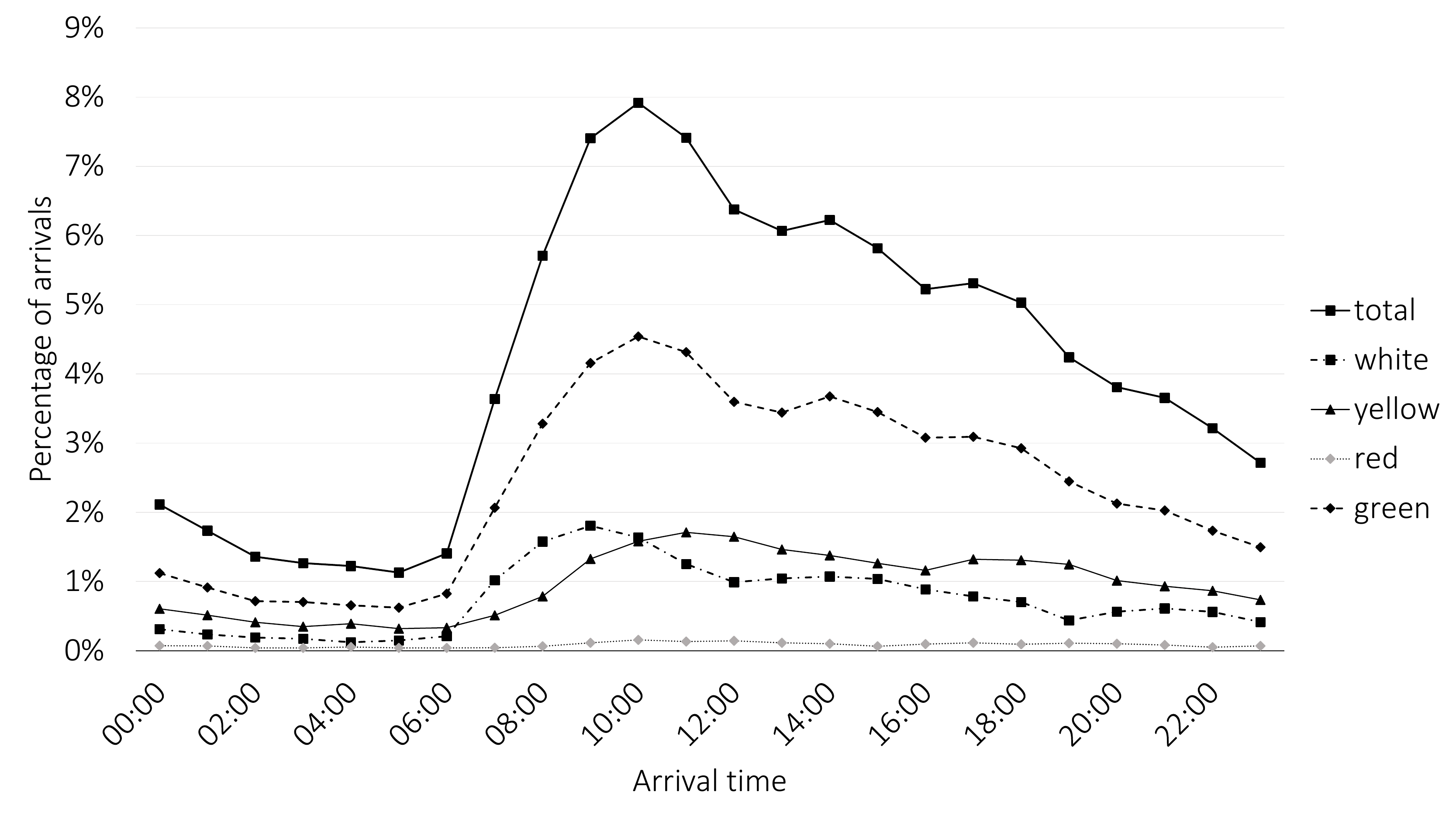}
	\caption{Arrival distribution per urgency code and time of day}
	\label{fig:arrival_dist}
\end{figure}

Figures \ref{fig:urgency_dist} and \ref{fig:arrival_dist} illustrate the profiles of the urgency code and the patients' arrival distributions per hour of the day, respectively. From these figures, it is possible to observe that the majority of patients entering the ED are from the green category and that most of them arrive at the ED between 8:00 and 12:00.

\begin{figure}[htb]
	\centering
	\includegraphics[width=0.75\textwidth]{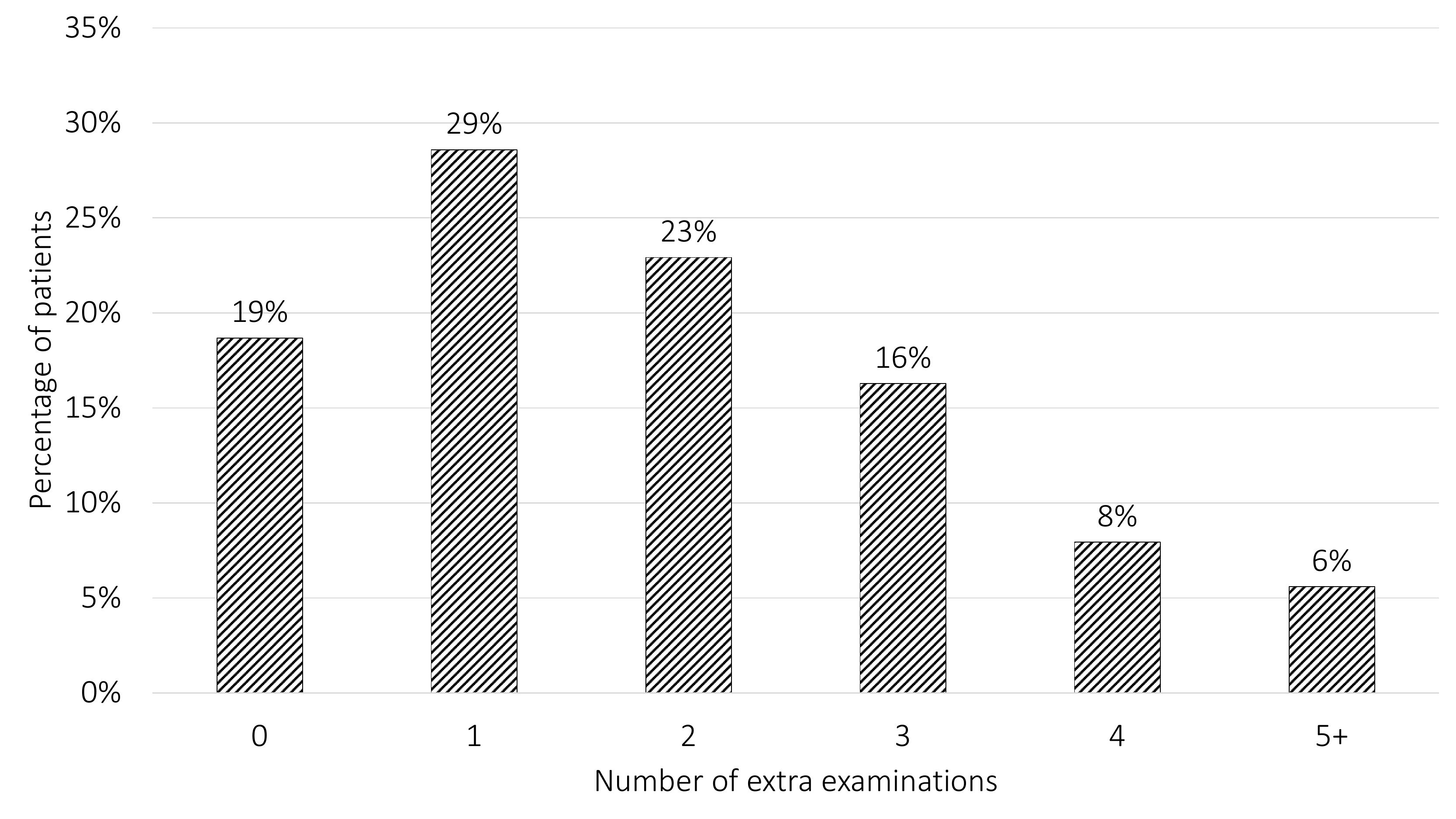}
	\caption{Extra examinations required apart from the first visit}
	\label{fig:nb_extra_exams}
\end{figure}

Concerning the first visits, they can be general ($79\%$), orthopaedic ($16\%$), or dermatological ($5\%$).
Regarding the additional exams required by ED patients in the past, it has been observed that for approximately $85\%$ of them, less than four extra examinations were required, as detailed in Figure \ref{fig:nb_extra_exams}.Besides, it has been identified that among all patients admitted to the ED, $57\%$ have been visited at the radiology area, and $54\%$ required laboratory exams. Altogether, these exams represented about $70\%$ of all exams performed.

As the laboratory examinations are performed in a building annexe to the main ED building, we investigated the total laboratory examination time more deeply. From our analysis, we observed that this time is mainly composed by (i) waiting time, (ii) effective examination time and (iii) miscellaneous times. Figure 6 shows how these different times impact on the total time and how they vary during the day. From Figure 6, it can be noted that the laboratory total time tends to be higher when the peak of patients arrivals is reached. From this figure, the presence of two peaks, at 7:00 and 19:00, can also be noted, demonstrating that the shift change impacts on the laboratory waiting times.
\begin{figure}[htb]
	\centering
	\includegraphics[width=0.75\textwidth]{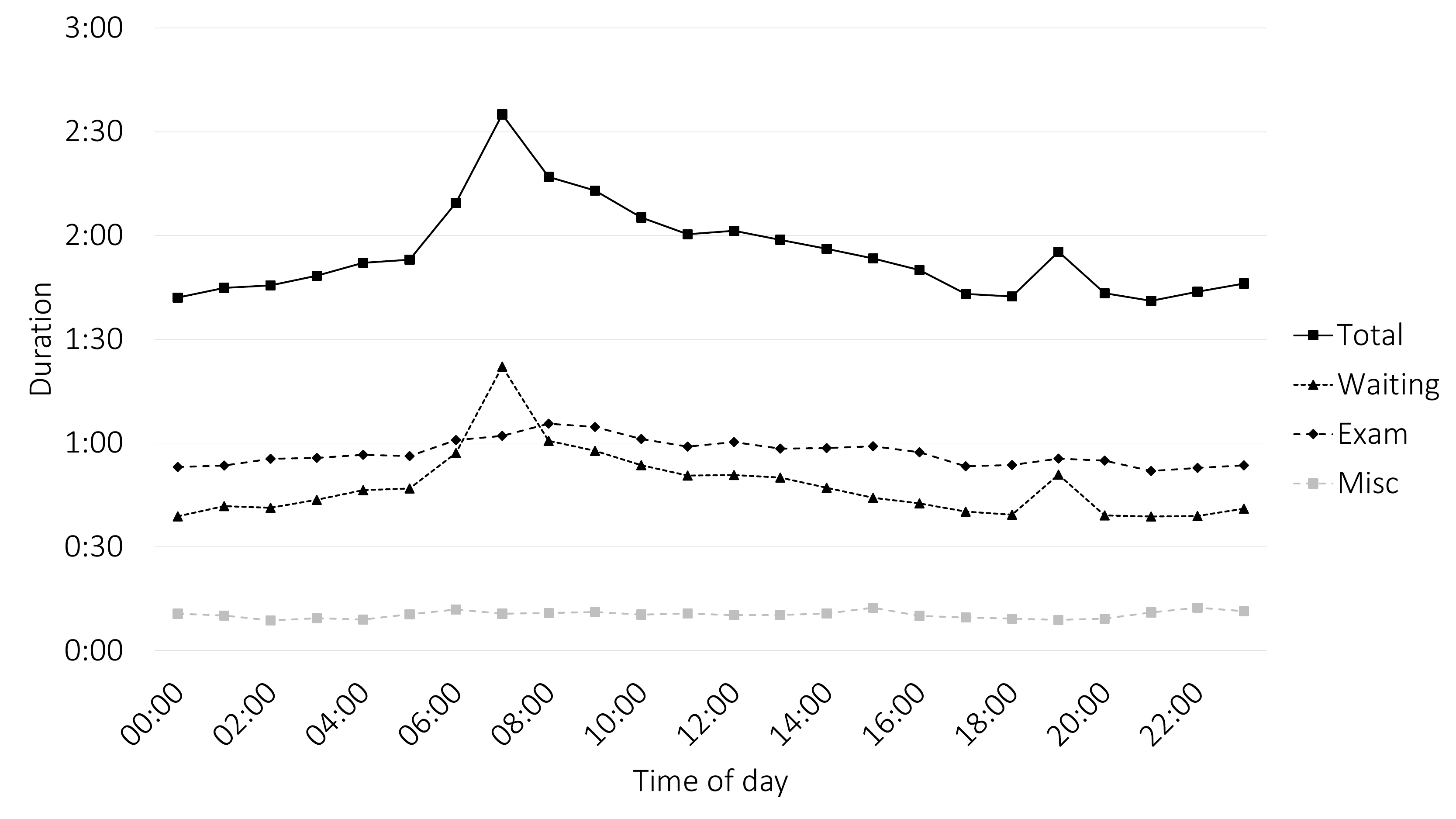}
	\caption{Laboratory service time composition}
	\label{fig:New_lab_service_time}
\end{figure}

\subsection{Model set-up and validation}
\label{subsec:validation_ED}

To better understand the ED, during the design thinking abstraction phase, we used the input data that we collected and analysed to build a DES model able to replicate the current ED setting. The developed model is structured taking into account the patient flow process depicted in Figure \ref{fig:flowchart} and all data collected and analysed in Section \ref{sec:data_ED}.

In the model, the patient arrivals are modelled using the inter-arrival data extracted from the database, and their triage category follows the data presented in Figure \ref{fig:arrival_dist}. The patient first visit type is also identified during the triage process and respects the proportion observed from the database. Then, patients are moved to the waiting room and are served following the first-in-first-out rule by category, i.e., red category patients have maximum priority, yellow category patients have priority over green and white ones, and so on. After the first visit, the patients can require laboratory examination with $54\%$ of probability, and, only after receiving their results, in case laboratory exams were required, they can follow their paths through the ED. Following the data shown in Figure \ref{fig:nb_extra_exams}, the patients can require extra examinations before being addressed to the last visit and then leave the system.

Once all crucial data have been considered and integrated into the developed DES model, the model needs to be validated. The validation phase is crucial when developing a simulation model and before using it. This importance is confirmed by several works in the literature dealing specifically with validation of simulation models (e.g., \citealt{Martis2006, Sargent2011}, just to cite some). This process is even more crucial when historical data is scarce, and assumptions made during the process need to be verified. The model validation phase has the goal of understanding how accurate the model is on simulating a real system.

\begin{figure}[htb]
	\centering
	\includegraphics[width=0.75\textwidth]{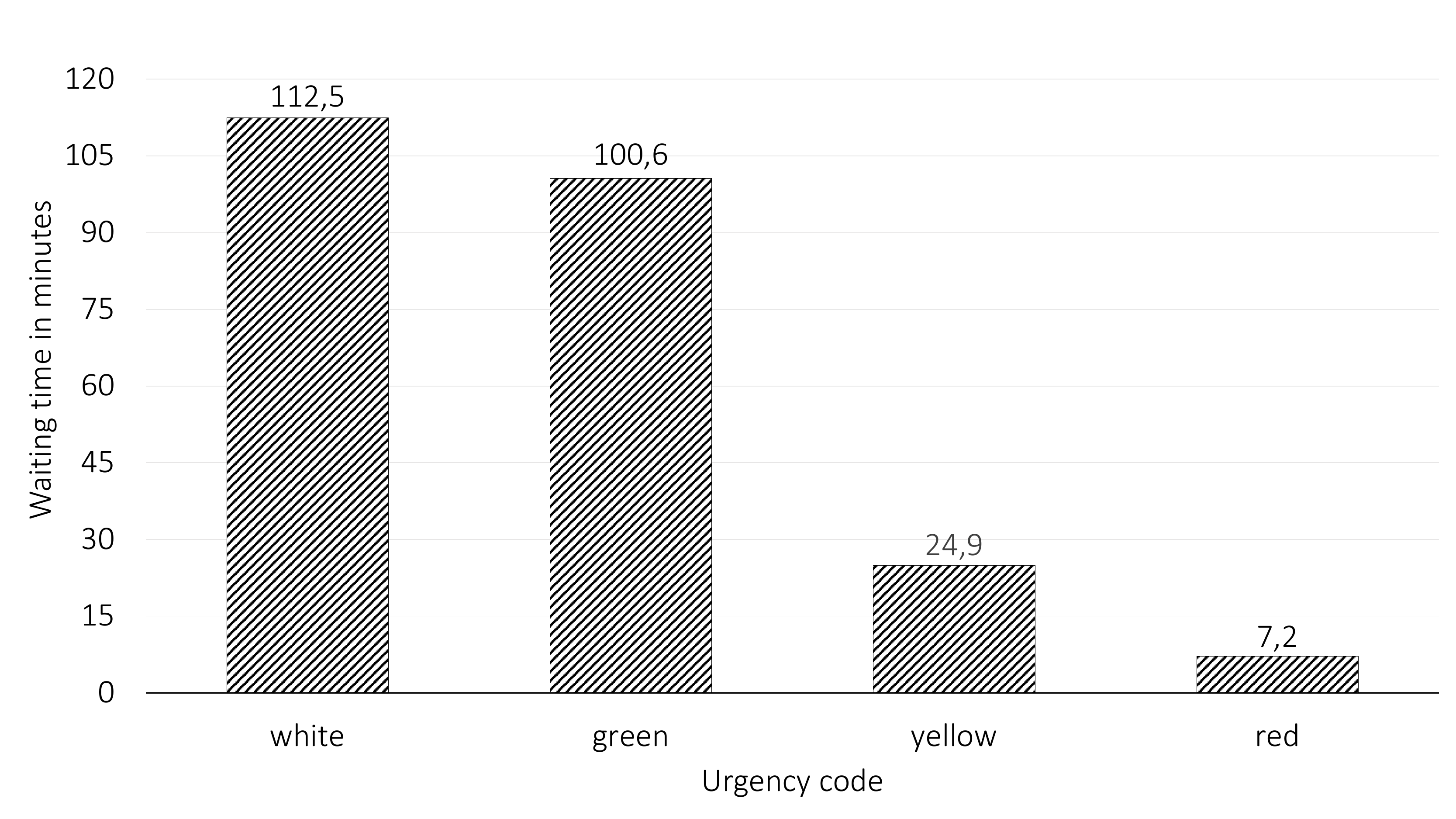}
	\caption{Average waiting times per urgency code}
	\label{fig:New_Waiting_times_b}
\end{figure}

\begin{figure}[htb]
	\centering
	\includegraphics[width=0.75\textwidth]{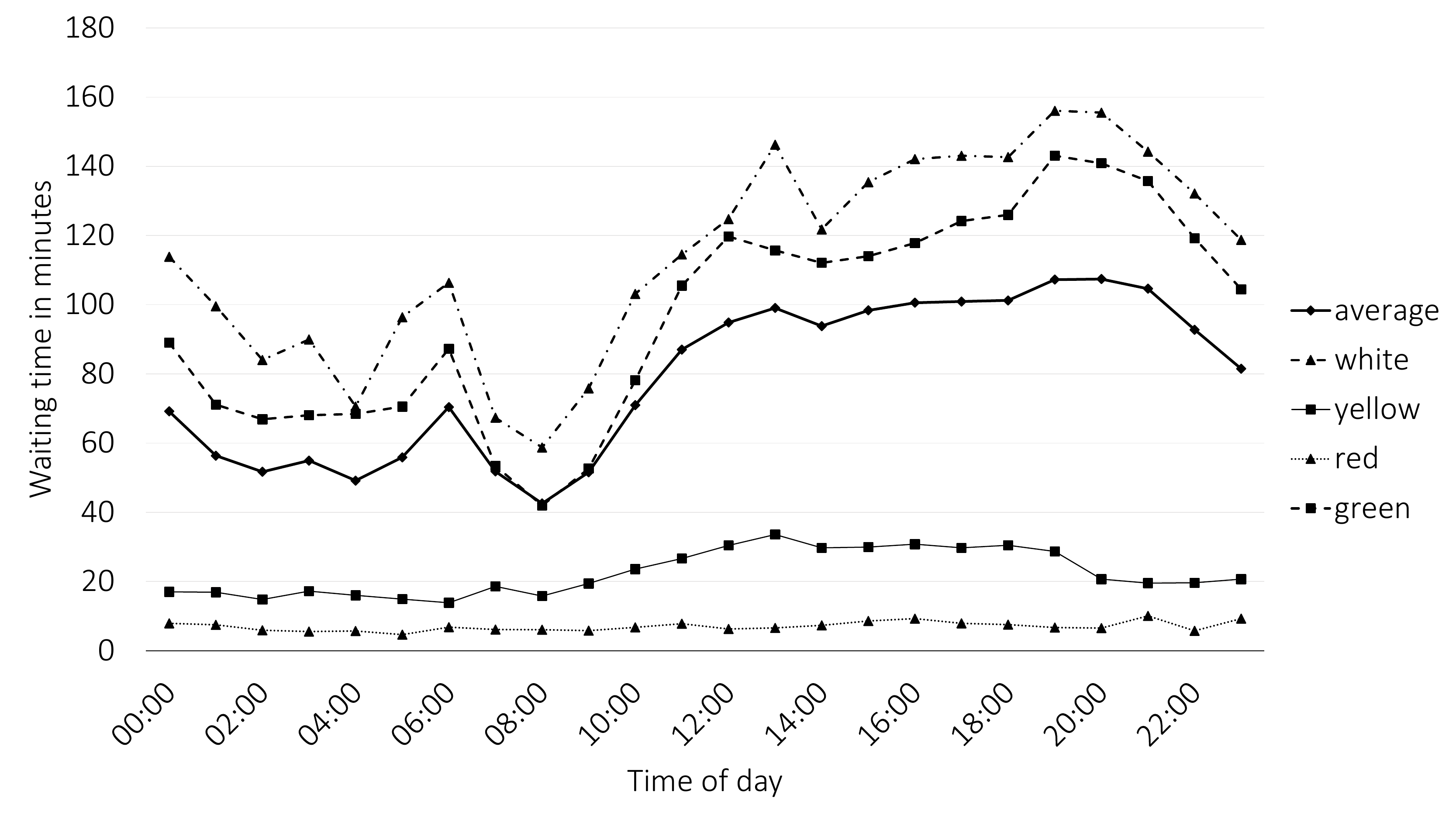}
	\caption{Average waiting times per urgency code and time of the day}
	\label{fig:New_Waiting_times}
\end{figure}

Our model validation phase is based on the consistent data retrieved from the ED database and is aligned with the one used by \cite{Aringhieri2010}. The output values from the simulation model were confronted with historical data. To do so, we identified the KPI of interest, namely the {\it length of stay} (LoS) and the {\it waiting time} (WT) for the first visit, that are correlated with the needs identified at the comprehension phase. These KPI are commonly used in simulation studies from the literature (see, e.g., \citealt{Marshall2005, Santibanez2009}).Our reference values for the LoS and WT have been obtained directly from the database and represent the average values for the available nine months of data. Figures \ref{fig:New_Waiting_times_b} and \ref{fig:New_Waiting_times} show the waiting times profile.

{In addition, we also considered outliers as a KPI. By outliers, we mean the percentage of patients who wait more than a given threshold time. This threshold value depends on the patients' urgency code and the ED internal, regional and national regulations. The reference threshold values have been obtained directly from the guidelines document by the Italian Ministry of Health  (\citealt{MinSalute2001b}).
	
The model has been validated by comparing the KPI from the historical data with those obtained by the simulation of the current setting. On the one hand, the historical results represent the average values for nine months of operations. On the other hand, the simulation results were obtained by running the simulation model ten times, where each run simulates one month of ED work. The obtained average results are presented in Table \ref{tab:validation}. Column {\it In} indicates the average number of patients arriving per day and columns {\it WT$_{1{st}}$} and {\it LoS} show the average waiting time for the first visit and the average length of stay (in minutes), respectively. Finally, columns {\it Outliers (\%)} present the percentage of patients, by priority, which exceeded the threshold time for waiting for the first visit.
\begin{table}[htbp]
	\centering
	\setlength{\tabcolsep}{0.8mm}
	\caption{Model validation results}
	{\begin{tabular}{lrrrrrHrrrrr}
			\toprule
			\multirow{2}{*}{Scenario} & & \multirow{2}{*}{In} & & \multirow{2}{*}{ {\it WT$_{1{st}}$} } & & \multirow{2}{*}{ {\it WT$_{\text{last}}$} } & & \multirow{2}{*}{LoS} & & \multicolumn{2}{c}{Outliers (\%)} \\
			\cmidrule{11-12}
			& & & & & & & & & & green & white \\
			\cmidrule{1-1}\cmidrule{3-12}
			Past & & 237.03 & & 83.27 & & - & & 206.43 & & 9.62 & 14.67 \\
			Simulation & & 238.23 & & 70.52 & & 54.94 & & 208.60 & & 3.88 & 25.47 \\
			\bottomrule
	\end{tabular}}%
	\label{tab:validation}%
\end{table}%
	
As can be seen in Table \ref{tab:validation}, the obtained values for {\it WT$_{1{st}}$} and {\it outliers} are not as accurate as the one obtained for the LoS. This fact is mainly due to the lack of quantitative data concerning service times and to the lack of an established queue rule in the ED. The values obtained for the LoS are, instead, very accurate. Based on this, the model has been approved by ED managers and staff involved in the study.
	
\subsection{Scenarios proposition}
\label{sec:scenarios_ED}
	
This section presents a set of scenarios that have been proposed to improve the ED performance. We developed seven main what-if scenarios, inspired by the literature and based on the needs, the weaknesses, and the possible intervention points identified. The major problems identified are the following ones: (1) the work shifts may not be fitted to the arrival pattern of the patients; (2) many patients wait a long time for the last visit before being dismissed; (3) the fixed priority rules for the visit may cause long waiting times for the less urgent patients; (4) for most of the patients, laboratory exams are required, but only at the end of the first visit; (5) many white urgency patients do not need any emergency service; (6) the work teams seem to be overcharged; (7) the blood sample transportation required for the laboratory exams is very inefficient and usually requires a significant amount of time.
	
To tackle these weaknesses, some possible goals have been established, namely: (1) adjust the team shifts to the demand; (2) increase the priority for the last visit over the first one; (3) dynamically change the patients’ queue priority, based on their WT; (4) identify and require laboratory exams at the triage process; (5) reduce the number of non-eligible patients that arrive at the ED; (6) test the possibility of using additional work teams; (7) improve the transportation system for the blood samples.

To attain the goals mentioned above, the following practical actions could be associated with them: (i) offset the starting and ending times for the team shifts; (ii) implementation of an alert system to support the dynamic priority rule; (iii) improving triage process by immediately dismissing non-eligible white urgency code patients and by requiring laboratory exams to a specific group of patients during this process; (iv) admit and train personnel for an additional work team dedicated to the last visit; and (v) implementing a more efficient transportation system.

Based on the discussion above, we defined the following parameters to characterise our proposed scenarios:
\begin{enumerate}
	\item {$t$: team shifts start (and finish) $t$ hours later than in the current setting;}
	\item $p$: if $p=1$ then the last visit has priority over the first one, otherwise, if $p=0$ the priority is as is in the current setting;
	\item $\tau_g$ and $\tau_w$: patients with green or white urgency codes can go to the head of the queue for the first visit if their current waiting time exceeds $\tau_g$ and $\tau_w$ minutes, respectively;
	\item $e$: $e\%$ of white priority code patients are not admitted to the ED, assuming that they could be directed to more appropriate facilities;
	\item $l$: $l\%$ of the laboratory exams could be required during the triage process;
	\item $a$: $a$ additional work teams are considered;
	\item $r$: the lead time for laboratory exams are reduced by $r$ minutes by avoiding long transportation times.
\end{enumerate}
\begin{table}[htbp]
	\centering
	\setlength{\tabcolsep}{0.8mm}
	\caption{Parameters}
	{\begin{tabular}{lrr}
			\toprule
			Parameter & & Chosen values \\
			\cmidrule{1-1}\cmidrule{3-3}
			$t$ & & {1, 2}\\
			$p$ & & {1}\\
			$\tau_g$ & & {60, 90, 120, 210} \\
			$\tau_w$ & & {120, 180, 210} \\
			$e$ & & {5, 10, 15, 20} \\
			$l$ & & {10, 15, 20, 50, 60, 75, 100} \\
			$a$ & & {1}\\
			$r$ & & {10, 15, 20, 25, 30} \\
			\bottomrule
	\end{tabular}}%
	\label{tab:parameters}%
\end{table}%
	
Table \ref{tab:parameters} shows the chosen parameter values. By combining them, we establish our notion of scenario. Hence, let us define $S = (t, p, \tau_g, \tau_w, e, l, a, r)$ as a generic scenario formed by a combination of these parameters. For example, a scenario $S = (-, -, 120, -, 5, -, -, 10)$ means that: the queue priority for green coded patients changes if their waiting time exceeds 120 minutes; $5\%$ of white urgency code patients are dismissed during the triage process, and the lead time for laboratory exams is $10$ minutes shorter. The ``$-$" sign states that the current configuration is not changed. Based on this observation, Table \ref{tab:scenarios_ED} presents the proposed scenarios that we evaluated.
Scenarios {\it A} considers a change in the personnel work shift while scenario {\it B} gives priority to the last visits over the first ones. Scenarios {\it C} are those proposing a new dynamic priority rule for the queue to the first visit. Scenarios {\it D} and {\it E} simulate an improvement in the triage process. The former implies requiring a percentage of laboratory exams during the triage process, while the latter seeks to reduce the number of non-eligible patients in the ED. In scenario {\it F} the addition of a new work team for the last visit is tested, and scenarios {\it G} simulate the reduction in the laboratory lead time. Finally, the last scenarios, labelled {\it Cb}, are formed by combining multiple types of scenarios from {\it A} to {\it G}.	
\begin{table}[htbp]
	\centering
	\setlength{\tabcolsep}{0.6mm}
	\caption{Proposed scenarios}
	{\begin{tabular}{ccccccccccccccccccccccc}
			\cmidrule{1-11}\cmidrule{13-23}    \multirow{2}[4]{*}{Id} & \multicolumn{10}{c}{Parameters}                                                 &       & \multirow{2}[4]{*}{Id} & \multicolumn{10}{c}{Parameters} \\
			\cmidrule{2-11}\cmidrule{14-23}          &       & $t$     & $p$     & $\tau_g$ & $\tau_w$ & $e$     & $l$     & $a$     & $r$     &       &       &       &       & $t$     & $p$     & $\tau_g$ & $\tau_w$ & $e$     & $l$     & $a$     & $r$     &  \\
			\cmidrule{1-11}\cmidrule{13-23}    A.1   & (     & 1     & --,   & --,   & --,   & --,   & --,   & --,   & --    & )     &       & F.1   & (     & --,   & --,   & --,   & --,   & --,   & --,   & 1     & --    & ) \\
			\cline{13-23}    A.2   & (     & 2     & --,   & --,   & --,   & --,   & --,   & --,   & --    & )     &       & G.1   & (     & --,   & --,   & --,   & --,   & --,   & --,   & --,   & 10    & ) \\
			\cline{1-11}    B.1   & (     & --,   & 1     & --,   & --,   & --,   & --,   & --,   & --    & )     &       & G.2   & (     & --,   & --,   & --,   & --,   & --,   & --,   & --,   & 15    & ) \\
			\cline{1-11}    C.1   & (     & --,   & --,   & 90,   & 180,  & --,   & --,   & --,   & --    & )     &       & G.3   & (     & --,   & --,   & --,   & --,   & --,   & --,   & --,   & 20    & ) \\
			C.2   & (     & --,   & --,   & 210,  & 210,  & --,   & --,   & --,   & --    & )     &       & G.4   & (     & --,   & --,   & --,   & --,   & --,   & --,   & --,   & 25    & ) \\
			C.3   & (     & --,   & --,   & 60,   & 120,  & --,   & --,   & --,   & --    & )     &       & G.5   & (     & --,   & --,   & --,   & --,   & --,   & --,   & --,   & 30    & ) \\
			\cline{13-23}    C.4   & (     & --,   & --,   & 90,   & --,   & --,   & --,   & --,   & --    & )     &       & Cb.1  & (     & --,   & --,   & 120,  & --,   & 10,   & 50,   & --,   & --    & ) \\
			C.5   & (     & --,   & --,   & 60,   & 180,  & --,   & --,   & --,   & --    & )     &       & Cb.2  & (     & --,   & --,   & 120,  & --,   & 10,   & 20,   & --,   & --    & ) \\
			C.6   & (     & --,   & --,   & 60,   & 210,  & --,   & --,   & --,   & --    & )     &       & Cb.3  & (     & --,   & --,   & 120,  & --,   & 10,   & 50,   & --,   & 30    & ) \\
			C.7   & (     & --,   & --,   & 120,  & --,   & --,   & --,   & --,   & --    & )     &       & Cb.4  & (     & --,   & --,   & 120,  & --,   & --,   & 50,   & --,   & --    & ) \\
			\cline{1-11}    D.1   & (     & --,   & --,   & --,   & --,   & --,   & 50,   & --,   & --    & )     &       & Cb.5  & (     & --,   & --,   & 90,   & --,   & 10,   & 50,   & --,   & --    & ) \\
			D.2   & (     & --,   & --,   & --,   & --,   & --,   & 60,   & --,   & --    & )     &       & Cb.6  & (     & --,   & --,   & --,   & --,   & --,   & 10,   & --,   & 15    & ) \\
			D.3   & (     & --,   & --,   & --,   & --,   & --,   & 75,   & --,   & --    & )     &       & Cb.7  & (     & --,   & --,   & --,   & --,   & --,   & 20,   & --,   & 15    & ) \\
			D.4   & (     & --,   & --,   & --,   & --,   & --,   & 100,  & --,   & --    & )     &       & Cb.8  & (     & --,   & --,   & --,   & --,   & --,   & 10,   & --,   & 20    & ) \\
			D.5   & (     & --,   & --,   & --,   & --,   & --,   & 10,   & --,   & --    & )     &       & Cb.9  & (     & --,   & --,   & --,   & --,   & --,   & 15,   & --,   & 20    & ) \\
			D.6   & (     & --,   & --,   & --,   & --,   & --,   & 15,   & --,   & --    & )     &       & Cb.10 & (     & --,   & --,   & --,   & --,   & --,   & 20,   & --,   & 20    & ) \\
			D.7   & (     & --,   & --,   & --,   & --,   & --,   & 20,   & --,   & --    & )     &       & Cb.11 & (     & --,   & --,   & --,   & --,   & --,   & 10,   & --,   & 30    & ) \\
			\cline{1-11}    E.1   & (     & --,   & --,   & --,   & --,   & 5,    & --,   & --,   & --    & )     &       & Cb.12 & (     & --,   & --,   & --,   & --,   & --,   & 15,   & --,   & 30    & ) \\
			E.2   & (     & --,   & --,   & --,   & --,   & 10,   & --,   & --,   & --    & )     &       & Cb.13 & (     & --,   & --,   & 120,  & --,   & 15,   & --,   & --,   & --    & ) \\
			E.3   & (     & --,   & --,   & --,   & --,   & 15,   & --,   & --,   & --    & )     &       & Cb.14 & (     & --,   & --,   & --,   & --,   & --,   & 50,   & --,   & 30    & ) \\
			E.4   & (     & --,   & --,   & --,   & --,   & 20,   & --,   & --,   & --    & )     &       & Cb.15 & (     & --,   & --,   & 120,  & --,   & 15,   & 50,   & --,   & 30    & ) \\
			\cmidrule{1-10}\cmidrule{13-23}
	\end{tabular}}%
	\label{tab:scenarios_ED}%
\end{table}%
	
\section{Scenario analytical evaluation}
\label{sec:experiments_ED}

In this section, we present and evaluate our proposition by simulating the scenarios presented in Section  \ref{sec:scenarios_ED}. The DES model was implemented using the software AnyLogic 8.1.0. The experiments were executed on a PC equipped with an Intel Core i7-7500U 2.70GHz processor and 12GB of RAM.

Each proposed scenario was simulated ten times, where each run simulated 30 days of working activity in the ED. We compare the average results of these runs with those obtained by the ten times 30 days simulations of the current ED setting in  \ref{tab:validation}. As comparison reference values, we used the same KPI previously described in Section \ref{subsec:validation_ED} (the LoS, the WT$_{\text{1st}}$, and number of outliers) to evaluate the scenarios proposed in Table 3. In addition, for the simulated results, we also show the average results for the WT for the last visit, referred to as WT$_{\text{last}}$.
	
Table \ref{tab:results_AllScenarios} summarises the results we obtained. The values in boldface indicate a significant KPI change by the referenced simulated scenario when compared to the current simulated setting.
The results obtained for scenarios $A$ indicate that offsetting the team shifts would contribute to reduce the average LoS of the patients, mainly due to a reduction in the waiting time for the last visit.
For the tested scenario $B$ (where the last visits have priority over the first ones), it can be noted that the average LoS is also reduced mainly due to the small waiting times for the last visit. However, the other KPI considered are worsened, especially the waiting time for the first visit and the percentage of outliers.
The obtained results show that the scenarios of type $C$ have a direct impact on the outliers indicator. This is expected because the queue priorities for patients with a long WT change when this value approximates the thresholds values.
The scenarios of type $D$ consider that a percentage of laboratory exams are required during the triage process. As this service is performed without requiring the presence of the patients, a saving time for waiting for the results is attained, as presented in Table \ref{tab:results_AllScenarios}}, thus impacting positively in the LoS indicator.
Concerning the results for scenarios of type $E$, the actions related to them act directly in the number of patients arriving at the ED, thus obtaining improving values for most of the considered KPI.
For scenario o $F$, an additional work team is considered and, similarly to the results for scenarios $E$, this allows improvements for all considered KPI.
It can be observed that for scenarios $G$ a consistent reduction in the LoS is achieved, as expected, because in these scenarios the laboratory results would be ready earlier than in the current setting.
In general, scenarios $D$ and $G$ act directly on the system bottleneck, i.e., the laboratory exams, and are thus very effective. It can be observed that, in both cases, the average LoS is reduced, which is a relevant result.
Finally, with regard to the results for the combined scenarios, it can be noticed that all of them were able to improve the average LoS. This behaviour is mainly due to the fact that these scenarios combine the best characteristics from scenarios of type $C$, $D$, $E$, and $G$. In particular, scenarios $Cb.3$ and $Cb.15$ present a reduction in the LoS of about $16\%$ and $19\%$, respectively.
\begin{table}[!htb]
	\centering
	\setlength{\tabcolsep}{1.5mm}
	\caption{Results obtained for the tested scenarios}
	{\begin{tabular}{lrrrrrrr}
			\toprule
			\multirow{2}{*}{Scenario} && \multirow{2}{*}{In} & \multirow{2}{*}{ {\it WT$_{\text{1st}}$} } & \multirow{2}{*}{ {\it WT$_{\text{last}}$} } & \multirow{2}{*}{LoS} & \multicolumn{2}{c}{Outliers (\%)} \\
			\cmidrule{7-8}
			&& & & & & green & white \\
			\cmidrule{1-1}\cmidrule{3-8}
			Curr. setting && \multirow{2}{*}{238.23} & \multirow{2}{*}{70.52} & \multirow{2}{*}{54.94} & \multirow{2}{*}{208.60} & \multirow{2}{*}{3.89} & \multirow{2}{*}{25.48} \\
			simulation \\
			\cmidrule{1-1}\cmidrule{3-8}
			A.1 && 236.93 & 68.67 & \textbf{45.34} & \textbf{198.42} & 3.48 &23.79\\
			A.2 && 237.35 & 71.61 & \textbf{38.58} & \textbf{195.37} & 3.30 & 26.71\\
			\cmidrule{1-1}\cmidrule{3-8}
			B.1 && 235.64 & \textbf{99.91} & \textbf{5.93} & \textbf{196.31} & \textbf{12.04} & \textbf{37.95} \\
			\cmidrule{1-1}\cmidrule{3-8}
			C.1 && 237.41 & 66.98 & 53.73 & 204.02 & 3.26 & 23.29 \\
			C.2 && 237.41 & 69.38 & 51.15 & 204.07 & 4.31 & {\bf 8.75} \\
			C.3 && 236.56 & 69.10 & 55.11 & 206.74 & 3.73 & 23.91 \\
			C.4 && 237.43 & 72.54 & 43.40 & {\bf 199.67} & {\bf 0.00} & 29.32 \\
			C.5 && 237.74 & 69.19 & 55.63 & 207.54 & {\bf 1.61} & 23.88 \\
			C.6 && 237.30 & 69.28 & 55.33 & 207.00 & {\bf 1.23} & 23.70 \\
			C.7 && 237.34 & 69.64 & 53.76 & {\bf 205.97} & {\bf 0.24} & {23.98} \\
			\cmidrule{1-1}\cmidrule{3-8}
			D.1 && 236.45 & 69.34 & 52.61 & {\bf 194.42} & 4.14 & 23.60 \\
			D.2 && 237.65 & 69.06 & 52.78 & {\bf 191.92} & 3.80 & 24.31 \\
			D.3 && 236.71 & 68.35 & 51.28 & {\bf 186.85} & 3.81 & 23.89 \\
			D.4 && 237.55 & 70.93 & 53.84 & {\bf 185.97} & 3.91 & 25.70 \\
			D.5 && 236.93 & 71.01 & 56.32 & 207.67 & 4.58 & 26.42 \\
			D.6 && 237.35 & 69.79 & 54.12 & {\bf 202.90} & 4.37 & 25.63 \\
			D.7 && 237.91 & 68.05 & 51.76 & {\bf 198.47} & 3.34 & 23.25 \\
			\cmidrule{1-1}\cmidrule{3-8}
			E.1 && 235.74 & 67.67 & 53.93 & {\bf 203.98} & 3.52 & 24.27 \\
			E.2 && 237.58 & \textbf{62.91} & {\bf 47.23} & {\bf 195.00} & 2.81 & \textbf{20.45} \\
			E.3 && 231.47 & {\bf 59.93} & {\bf 45.19} & {\bf 190.78} & 2.77 & \textbf{20.75} \\
			E.4 && 228.00 & {\bf 53.88} & {\bf 40.25} & {\bf 182.05} & \textbf{1.97} & \textbf{16.73} \\
			\cmidrule{1-1}\cmidrule{3-8}
			F.1 && 236.36 & \textbf{39.91} & \textbf{24.36} & \textbf{154.97} & \textbf{0.37} & \textbf{7.17} \\
			\cmidrule{1-1}\cmidrule{3-8}
			G.1 && 237.19 & 69.31 & 54.43 & {\bf 202.25} & 3.77 & 23.76 \\
			G.2 && 236.73 & 68.39 & 53.74 & {\bf 199.30} & 3.56 & 24.37 \\
			G.3 && 236.44 & 68.26 & 53.34 & {\bf 197.83} & 3.63 & 23.22 \\
			G.4 && 236.65 & 69.08 & 53.09 & {\bf 196.80} & 3.24 & 24.47 \\
			G.5 && 237.65 & 69.37 & 54.03 & {\bf 195.42} & 4.12 & 24.62 \\
			\cmidrule{1-1}\cmidrule{3-8}
			Cb.1 && 234.79 & \textbf{65.38} & 51.49 & \textbf{191.07} & \textbf{0.04} & 22.27 \\
			Cb.2 && 232.98 & \textbf{63.22} & \textbf{47.81} & \textbf{192.54} & \textbf{0.13} & 22.10 \\
			Cb.3 && 233.35 & \textbf{61.38} & \textbf{45.72} & \textbf{174.29} & \textbf{0.08} & 20.10 \\
			Cb.4 && 237.96 & 72.13 & 53.14 & \textbf{197.02} & \textbf{0.14} & 26.26 \\
			Cb.5 && 237.28 & 75.92 & \textbf{45.11} & \textbf{194.79} & \textbf{0.02} & \textbf{32.06} \\
			Cb.6 && 235.70 & 67.66 & 53.50 & {\bf 196.74} & 3.82 & 23.02 \\
			Cb.7 && 237.08 & 68.93 & 53.61 & {\bf 195.86} & 3.64 & 24.97 \\
			Cb.8 && 236.35 & 69.50 & 52.04 & {\bf 195.50} & 4.49 & 24.28 \\
			Cb.9 && 237.68 & 69.94 & 52.41 & {\bf 195.33} & 4.02 & 24.99 \\
			Cb.10 && 236.62 & 69.20 & 52.68 & {\bf 194.46} & 4.23 & 23.79 \\
			Cb.11 && 236.45 & 69.78 & 51.58 & {\bf 192.45} & 3.95 & 24.23 \\
			Cb.12 && 238.40 & 69.15 & 53.92 & {\bf 192.68} & 3.73 & 24.38 \\
			Cb.13 && 230.39 & {\bf 56.55} & {\bf 44.46} & {\bf 187.19} & {\bf 0.08} & {\bf 18.18} \\
			Cb.14 && 236.83 & 67.97 & 53.17 & {\bf 185.73} & 3.65 & 24.24 \\
			Cb.15 && 230.62 & {\bf 57.56} & {\bf 41.69} & {\bf 168.37} & {\bf 0.12} & {\bf 18.84} \\
			\bottomrule
	\end{tabular}}%
	\label{tab:results_AllScenarios}%
\end{table}
	
\section{Scenario implementation evaluation and final decision}
\label{sec:experiments_ED_eval}

As previously mentioned, every three months, for the whole duration of the project, there has been a three hours meeting where the working team discussed the proposed scenarios with the top management. The structure of the meeting was designed in order to support factual discussion. The opening was a detailed presentation of the scenarios with an in-depth explanation of the simulation, the data collecting process, and the validation with the tests performed. This first part was followed by an in-depth Q\&A where head physicians and hospital top management were clarifying any doubts. The last part of the meeting was an open discussion with the ED team members. As a result of these meetings, we report that the choices suggested in Section \ref{sec:experiments_ED} did not reflect the final choices made by the group appointed to decide (doctors, nurses, engineers, and designers).
	
Indeed, scenarios $A$ and $F$ have not been accepted by the group despite the positive results of the simulation. Scenario $A$ presented practical difficulties in changing the team shifts, and hiring a new team was not a viable option from the top management. Scenario $B$ imposed a fixed rule of re-evaluation of patients to all doctors that at the moment can freely decide when to re-evaluate patients. This scenario has been evaluated as a promising alternative to speed-up the dismissing process. However, the team evaluated the difficulties of imposing a formal rule to give priority to the re-evaluation visit over the first visit and the possibility to spark internal conflicts defining a reference time. For these reasons, the team decided to reject this scenario. Scenario $C$ has been well accepted by the ED staff, as it represents an alternative for managing the waiting room by applying effective and straightforward rules. Scenarios $D$ and $E$ have been accepted as a promising option to reduce overcrowding. Nevertheless, implementing Scenarios $C$, $D$, and $E$ would require an in-depth training process and, for scenario $D$, a supplemental investigation to identify which exams would be required based on patients’ symptoms. That is why the team decided to postpone the implementation of scenarios $C$, $D$, and $E$ to a later date, to understand whether other implementation options would require less training and impact on the organisation. Finally, scenario $G$ deals with a faster sub-process of laboratory exams. In this scenario, the main concern is the speeding up of the subprocess from the moment in which the doctor requests a laboratory exam to the moment the laboratory receives the test tube. That scenario has been well accepted by ED managers and considered as the\textquoteleft easiest\textquoteright to be implemented.
	
After the top management approval, the team designed and planned practical actions for scenario $G$ implementation. Three organisational prototypes were hypothesised and simulated: (i) more frequent deliveries of test tubes to the central laboratory by the actual transport supplier; (ii) internal aid nurses dedicated to the transport of test tubes; (iii) pneumatic post design. Those three prototypes responded to different G scenarios, as more frequent deliveries could speed up the process (i) from $90$ to $30$ minutes, internal aid nurses involved could push that scenario to (ii) $25$ minutes and the pneumatic post to (iii) $10$ minutes. The innovation office made a first assessment of the investments required by the three prototypes, and with surprise, the pneumatic post had the highest ratio of cost savings/investments. The top management decided to extend the pneumatic post to the whole building where the ED is located, and they approved the rules to access the pneumatic post (hours of the day and urgency of exams). The top management allocated the budget for the investment in a pneumatic post system across the hospital, connecting the ED building and the exam lab building. Thus, the pneumatic post requirement was designed by the innovation and technical office of the hospital. A public announcement to build the pneumatic post system in the hospital was launched only a few months later. With precise data coming from the simulation, the top management and all the stakeholders had a clear picture of the increase of service quality for patients and staff, and the savings connected. This clarity helped to exert the right amount of pressure on the whole organisation so that the project could move fast, and the pneumatic post is now working at the hospital
	
The project thus answered to the top management kick off constraints, that required an implemented and working choice in less than two years from the kick-off of the project.
	
\section{Concluding remarks}
\label{sec:concl_ED}
	
The case study shows that organisational decisions driven by a pure simulation approach differ from the organisational decisions driven by a mixed approach simulation and design-driven. In the case study, we used a simulation modelling tool integrated into a design process with the primary goal of implementing a working solution in one year and a half from the beginning of the project. Improvements were measured by the reduction of patients waiting time and length of stay. The benefits of using simulation tools to model different solutions utilising what-if scenarios are recognised in literature, as well as the fact that it is essential to involve different stakeholders in their definition, analysis, and evaluation (\citealt{tako2015}). However, the embedding of the simulation tool into a design process leads to a different approach that couples a prototyping and testing approach to the simulation approach. In our research, we exploited the main strengths of these techniques by integrating the use of rigorous and extensive simulation modelling into a design thinking process. Indeed, this case study shows that not only the process but also the final decisions were different.
	
As future research direction, we plan to follow the implementation of the pneumatic post plan, to check if simulation results we obtained are consistent with the real implementation and extend the use of integrated simulation and design thinking approach to other real-world case studies. We also plan to tackle the problem of scheduling patients to doctors from an optimisation point of view. In this sense, the consideration of dynamic and stochastic versions of the problem, applied to real ED cases, seems to be an interesting and relevant future research direction to be followed.
		
\section*{Acknowledgements}

This research was funded by the CNPq - Conselho Nacional dae Desenvolvimento Cient{\'i}fico e Tecnol{\'o}gico, Brazil, grant No. 234814/2014-4 and by University of Modena and Reggio Emilia, grant FAR 2018 Analysis and optimisation of healthcare and pharmaceutical logistic processes. Their support is gratefully acknowledged.
		
\bibliographystyle{mmsbib}
\bibliography{ref}

\end{document}